\documentclass[namedreferences]{solarphysics}
\usepackage[optionalrh]{spr-sola-addons} % For Solar Physics
\usepackage{epsfig}          % For eps figures, old commands
\usepackage{graphicx}        % For eps figures, newer & more powerfull
\usepackage{natbib}         % For citations: redefine \cite commands
\usepackage{amssymb}        % useful mathematical symbols
\usepackage{color}           % For color text: \color command
\usepackage[T1]{fontenc}        % For typing Polish ogonek symbols (\k{} - A. Kepa co-author name)
\newcommand{\fex}{{Fe~{\sc x}}}

\newcommand{\fexiv}{{Fe~{\sc xiv}}}

\begin{document}

\begin{article}
\begin{opening}

\title{A Search for High-Frequency Coronal Brightness Variations in the 21 August 2017 Total Solar Eclipse}

\author{P.~\surname{Rudawy}$^{1}$ \sep K.~\surname{Radziszewski}$^{1}$ \sep A.~\surname{Berlicki}$^{1,4}$ \sep K.J.H.~\surname{Phillips\orcid{0000-0002-3790-990X}}$^{2}$ \sep D.B.~\surname{Jess}$^{3}$\,\sep P.H.~\surname{Keys}$^{3}$ \sep F.P.~\surname{Keenan}$^{3}$}

\runningauthor{P. Rudawy {\it et al.}}
\runningtitle{High-Speed Coronal Brightness Variations}

\institute{$^{1}$ Astronomical Institute, University of Wroc{\l}aw, 51-622 Wroc{\l}aw, ul. Kopernika 11, Poland email: \url{rudawy@astro.uni.wroc.pl; radziszewski@astro.uni.wroc.pl; berlicki@astro.uni.wroc.pl}  \\
$^{2}$ Earth Sciences Dept., Natural History Museum, Cromwell Road, London SW7 5BD, UK email: \url{kennethjhphillips@yahoo.com} \\
$^{3}$ Astrophysics Research Centre, School of Mathematics and Physics, Queen's University Belfast, Belfast BT7 1NN, Northern Ireland, UK email: \url{d.jess@qub.ac.uk; p.keys@qub.ac.uk; f.keenan@qub.ac.uk}  \\
$^{4}$ Astronomical Institute, Czech Academy of Sciences, 25165 Ondrejov, Czech Republic  \\ }

\date{Received               ; accepted        }

%%%%%%%%%%%%%%%%%%%%%%%%%%%%%%%%%%%%%%%%%%%%%%%%%%%%%%%%%%%%%%%%%%%%%%%%%%%%%%%%%%%%%%%%%%%%%%%%%%%%%
% Abstract
\begin{abstract}
We report on a search for short-period intensity variations in the green-line (\fexiv\ 530.3\,nm) emission from the solar corona during the 21 August 2017 total eclipse viewed from Idaho in the United States. Our experiment was performed with a much more sensitive detection system, and with better spatial resolution, than on previous occasions (1999 and 2001 eclipses), allowing fine details of quiet coronal loops and an active-region loop system to be seen. A guided 200-mm-aperture Schmidt--Cassegrain telescope was used with a state-of-the-art CCD camera having 16-bit intensity discrimination and a field-of-view (\mbox{$0.43^{\circ} \times 0.43^{\circ}$}) that encompassed approximately one third of the visible corona. The camera pixel size was 1.55\,arcseconds, while the seeing during the eclipse enabled features of $\approx 2$\,arcseconds (1450\,km on the Sun) to be resolved. A total of 429 images were recorded during a 122.9\,second portion of the totality at a frame rate of 3.49\,s$^{-1}$. In the analysis, we searched particularly for short-period intensity oscillations and travelling waves, since theory predicts fast-mode magneto-hydrodynamic (MHD) waves with short periods may be important in quiet coronal and active-region heating. Allowing first for various instrumental and photometric effects, we used a wavelet technique to search for periodicities in some $404,000$ pixels in the frequency range $0.5 - 1.6$\,Hz (periods 2\,second to 0.6\,second). We also searched for travelling waves along some 65 coronal structures. However, we found no statistically significant evidence in either. This negative result considerably refines the limit that we obtained from our previous analyses, and it indicates that future searches for short-period coronal waves may be better directed towards Doppler shifts as well as intensity oscillations.
\end{abstract}
\end{opening}

%\maketitle

%%%%%%%%%%%%%%%%%%%%%%%%%%%%%%%%%%%%%%%%%%%%%%%%%%%%%%%%%%%%%%%%%%%%%%%%%%%%%%%%%%%%%%%%%%%%%%%%%%%%%
% Introduction
\section{Introduction}

Despite many years of theoretical and observational investigations, the basic physical mechanism (or group of mechanisms) responsible for the global coronal heating remains the subject of much debate \citep[see, {\it e.g.},][and references therein]{2015RSPTA.37340269D, 2017ApJ...849...46V}. However, the  most widely recognized candidates for the dominant heating mechanism are a direct current (DC) type and alternating current (AC) type. The DC mechanism is based on numerous small reconnections of the highly filamentary coronal magnetic field, giving rise to nanoflares \citep{1974ApJ...190..457L, 1988ApJ...330..474P, 2015ApJ...805...47P, 2017ApJ...837..108P}. Such local reconnections could be triggered by the random shuffling of flux-tube footpoints, forced by macroscopic motions of dense plasma in the convection layer and photosphere \citep{1974ApJ...190..457L, 1988ApJ...330..474P}. The AC mechanism is based on numerous small-scale episodes of local energy dissipations carried to the corona by Alfv\'{e}n and other magnetohydrodynamic (MHD) waves due to ion viscosity or electrical resistivity \citep{1982ApJ...254..806H, 2001MNRAS.326..428W, 2006SoPh..234...41K, 2012RSPTA.370.3193D, 2018MNRAS.476.3328M}. MHD waves possibly dissipate energy in the loop-like magnetic structures over a range of spatial scales, although these may be smaller than the resolution limit of present-day instruments \citep{1994ApJ...435..482P, 1997ApJ...478..799C, 1999ApJ...514..493K, 2015SSRv..190..103J}.

Generally, the efficiencies and occurrence of both mechanisms are defined by local plasma parameters, field topology, and the evolution of numerous small flux tubes filling the solar corona. Individual nanoflares are hypothesized to dissipate energies of the order of 10$^{17}$\,J per event \citep{1988ApJ...330..474P, 2008ApJ...683..516S}, which are below the detection limits of current solar instrumentation. \citet{1991SoPh..133..357H} pointed out that the observed energy distribution of very small flares, which has a power-law form ($E^{-n}$, $E = $ flare total energy), indicates that nanoflares must have a power-law index $n> 2$ to be significant for coronal heating. Observations \citep{1998ApJ...501L.213K, 2000ApJ...529..554P, 2004psci.book.....A} do indeed indicate this to be true, although the estimated total energy flux released by the nanoflares appears to be about a factor of three smaller than that required (300\,J\,m$^{-2}$\,s$^{-1}$) for the heating of the quiet corona.

Numerous searches for intensity or velocity oscillations in the solar corona at low frequencies have been made. Wave heating may be an important contributor or even dominate in some regions of open magnetic fields, where magnetic reconnections cause particle acceleration rather than heating of the plasma volume \citep{2004A&A...416.1179R}. The low-frequency ($\approx$0.002\,--\,0.01\,Hz) Alfv\'{e}n and Alfv\'{e}nic waves observed by \citet{2009Sci...323.1582J} and \citet{2007Sci...318.1574D}, respectively, have an energy flux of at least 100\,J\,m$^{-2}$\,s$^{-1}$, sufficient to power the solar wind and constituting a significant part of the solar corona energy budget \citep{2007SoPh..243....3K, 2008SoPh..249..167T}. Tomczyk and co-workers \citep{2007Sci...317.1192T, 2009ApJ...697.1384T}, using the ground-based {\it Coronal Multi-Channel Polarimeter} (CoMP) instrument on a coronagraph, detected low-frequency ($\approx 0.003$\,Hz) velocity waves propagating along active-region magnetic loops. However, the energy flux transported by these oscillations was estimated to be only 0.01\,J\,m$^{-2}$\,s$^{-1}$ \citep{2007Sci...317.1192T}. Low-frequency ($\approx$\,0.002\,--\,0.006\,Hz) transverse waves have also been observed in the corona by \citet{2013A&A...556A.124T} and \citet{2015NatCo...6E7813M}. In addition, eclipse and coronagraph observations \citep[{\it e.g.} ][]{1994A&A...281..249K, 1999SoPh..188...89C} have revealed significant oscillations, but in the frequency range 0.003\,--\,0.014\,Hz, which are (as with the Tomczyk {\em et al.} waves) ineffective in coronal heating.

Theoretical work has shown that high-frequency MHD waves (with frequencies $> 0.5$\,Hz) could effectively transport energy into the solar corona \citep[{\it e.g.} ][]{1994ApJ...435..482P}. The energy dissipated might give rise to periodic or quasi-periodic intensity variations, in particular in coronal visible-light emission lines such as the green line (\fexiv\ 530.3\,nm) or red line (\fex\ 637.4\,nm). Attempts to detect such variations have been summarized by \citet{2004A&A...416.1179R,2010SoPh..267..305R} and \citet{2009Natur.459..789P}. Briefly, the evidence remains equivocal, with the 2006 total eclipse observations of \citet{2009SoPh..260..125S} of intensity variations in coronal loops at frequencies between 0.037\,--\,0.05\,Hz in both the green and red lines perhaps providing the strongest detection to date.

Our observations with the {\it Solar Eclipse Coronal Imaging System} (SECIS) for eclipses in 1999 and 2001 have been analyzed in detail by \citet{2004A&A...416.1179R}, \citet{2001MNRAS.326..428W}, and \citet{2010SoPh..267..305R}. \citet{2004A&A...416.1179R}, using wavelet analysis, found numerous maxima with $> 4\sigma$ significance in the 0.1\,--\,1.0\,Hz power spectra of green-line observations obtained for the 1999 eclipse observed in Bulgaria, although stricter statistical tests cast doubt on their reality. Nevertheless, an extended wavelet analysis of the dataset showed some evidence of a travelling wave with frequency 0.16\,Hz moving along a small active-region loop \citep{2001MNRAS.326..428W}. \citet{2010SoPh..267..305R} analyzed an improved data set obtained during the 2001 eclipse observed from Zambia. Some 11,000 individual locations were examined using both classical Fourier and wavelet software; no statistically significant period, in the range 0.06\,--\,10\,Hz, was found among these locations. However, the CCD cameras employed in both the 1999 and 2001 eclipses, although state-of-the-art when manufactured in 1997, had effectively only a ten-bit capacity giving a dynamic range of only 1000:1. Furthermore, the spatial sampling, determined chiefly by the pixel size, was only \mbox{$4 \times 4$}\,arcseconds.

Although much more sophisticated space instrumentation is now available than was the case for the 1999 and 2001 eclipses, it remains true that space-based imaging cadences \citep[{\em e.g.} those of the {\it Atmospheric Imaging Assembly} (AIA) on board the {\it Solar Dynamics Observatory}:][]{2012SoPh..275...17L} are still too low to search for high-frequency ($\approx $1\,Hz) intensity variations. Hence there is a need to obtain high-quality, high-cadence imaging data with ground-based instruments. In addition, the solar corona is far better observed from the ground during total solar eclipses than with a ground-based coronagraph, since  for the latter there remains significant atmospheric scattering of light from the solar disk.

Here we report on the results of a search for high-frequency brightness variations in the \fexiv\ 530.3\,nm coronal green line, using data collected during the 21 August 2017 total solar eclipse by a team from Queen's University Belfast (UK), the University of Wroc{\l}aw (Poland), and the Natural History Museum in London (UK). In the following, Section\,2 describes in detail the instrumental set-up used for the observations, and Section\,3 the observing site. Criteria adopted for the selection of the target region are summarized in Section\,4, observation procedures in Section\,5, corrections for motions and coalignment of the images in Section\,6, and the photometric analysis of the data in Section\,7. Results and conclusions are given in Section\,8.

%--------------------------------------------------------------------
% Section 2
\section{Instrumentation}

Versions of the instrumentation, based on the original SECIS design, have been used by us during the total solar eclipses in 1999 and 2001 to search for brightness oscillations in the corona \citep{2000SoPh..193..259P, 2001MNRAS.326..428W, 2010SoPh..267..305R}. The instrument used for the 21\,August 2017 eclipse consisted of three main components: i) a 200 mm f/10 Celestron C8S XLT Schmidt-Cassegrain telescope to provide an image of the eclipsed Sun; ii) the SECIS optical box forming two images of the solar corona in white-light and the \fexiv\ 530.3\,nm green coronal line; iii) a pair of fast-frame Andor CCD cameras. All components were connected by a rigid optical bench installed on a SkyWatcher EQ-8 drive mounted on top of a heavy tripod. The complete instrument was tested before the expedition with night-time observations of the full Moon at the Astronomical Institute in Wroc{\l}aw.

An image of the solar corona, formed by the telescope, was projected into the SECIS optical box, which consisted of a collimating lens, a pellicle beam splitter, narrow-band (green-line) and neutral-density filters, and two output lenses (see, {\em e.g.}, \citet{2000SoPh..193..259P}; \citet{2010SoPh..267..305R} for details). The beam splitter has a 90\,\% transmission in the direct-line path and 10\,\% in the reflected path. The direct-line beam passed through a narrow-band filter centred on the \fexiv\ 530.3\,nm green line, and it was then focused on to an Andor iXon3 885 CCD 16-bit monochrome camera. The narrow-band interference filter centred on 530\,nm was prepared by Barr Associates, Inc., and it was selected as the best from a sample of three, using transmission measurements taken in March 2017 with a spectrophotometer at the Technical University of Wroc{\l}aw. The maximum filter transmittance was measured to be at a wavelength of 529.95 nm, and the bandpass width (FWHM) of 0.43 nm; this bandpass includes the laboratory measured (air) wavelength of \cite{2013ApJ...776..121S}. The Andor camera has \mbox{$1004 \times 1002$} square pixels of dimensions \mbox{$8 \times 8$\,$\mu$m$^2$} and a quantum efficiency of about 55\,\% at the green-line wavelength. The effective spatial sampling in the green-line channel was 1.55\,arcseconds per pixel (1140\,km on the Sun), yielding a field-of-view (FoV) of \mbox{$0.43^{\circ} \times 0.43^{\circ}$}, or approximately one third of the visible corona. The effective time cadence was 0.28651\,second, giving a frame rate of 3.49\,images\,s$^{-1}$.

The reflected-path beam of the system recorded white-light images, which formed a ``control'' for the analysis of the green-line images. This beam passed through a neutral-density filter with transmission 6.3\,\%, and images of the white-light corona were recorded with an Andor Zyla 5.5 sCMOS 16-bit monochrome camera, consisting of \mbox{$2560 \times 2160$} square pixels with dimensions \mbox{$6.5 \times 6.5$\,$\mu$m$^2$}. The spatial dimensions in the white-light channel were 1.27\,arcsecond per pixel (927\,km), yielding an FoV of 0.90$^{\circ}$ $\times$ 0.76$^{\circ}$. Thus, a larger portion of the corona was imaged as compared with the green-line images. Optical components of both the direct-line and reflected paths were installed in a rigid, light-tight aluminium box, and both CCD cameras were fixed directly to the box. The spatial resolution of the white-light channel, which is nominally marginally better than that of the green-line channel, was in fact comparable, as there was slight degradation due to  atmospheric seeing.

The instrument used was much improved over the instruments used by us in the 1999 and 2001 eclipses, which had only a 1000:1 dynamic intensity range and spatial resolution of 4\,arcseconds per pixel. The lower frame rate of our images led to intensity levels of the emitting structures that were higher by a factor of over 50 than those obtained with the 2001 eclipse version of our instrument.

% Section 3
\section{Observing Site}

The instruments were installed on a flat, grassy meadow in the Park Creek Ranch, located about 10\,km north of Stanley, a small settlement in the mountain region of Idaho, USA, at \mbox{W$114^{\circ}56^{\prime}24^{{\prime}{\prime}}$}; \mbox{N$44^{\circ}13^{\prime}12^{{\prime}{\prime}}$}, 1906\,m above sea level. This observing site was pre-selected as a region having a high probability of clear sky, and with easy access and convenient living quarters. The Park Creek Ranch occupies the bottom of a very wide and flat valley in the Rocky Mountains, near Salmon River. Apart from a few trees and bushes, there were no local obstacles in front of the instrument, and the horizon was defined by the distant Rocky Mountains. The elevation of the Sun during the totality (at 17:29\,UT or 11:29\,local time) at the observing site was \mbox{$46^{\circ}48^{\prime}$}, with azimuth \mbox{$128^{\circ}06^{\prime}$}.

%--------------------------------------------------------------------
% Section 4
\section{Target Selection}

\begin{figure}
   \centerline{\includegraphics[width=9cm]{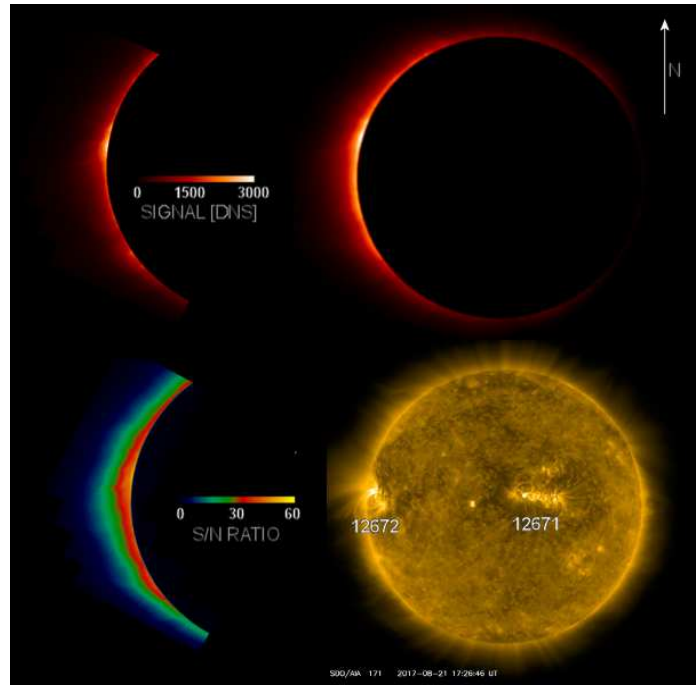}}
      \caption{Upper panels: (left) the eastern part of the corona seen in our \fexiv\ 530.3\,nm images; (right) the white-light corona as viewed seven seconds after eclipse second contact at 17:28:23\,UT. The contrast of the green-line image is enhanced to better display the filamentary coronal structures. Images are rotated to match the SDO/AIA image (North at the top, East on the left). Lower panels: (left) signal-to-noise ratio corresponding to the green-line image in the panel above; (right) SDO/AIA image taken at 17:26:46\,UT. NOAA active-region numbers are shown in the SDO/AIA image.}
      \label{figure1}
  \end{figure}

The level of solar activity on 21 August 2017 was generally low. Only two active regions were present on the solar disk on the day of the eclipse: an extensive active region (NOAA 12671) of $\beta \gamma$ magnetic type, located near disk centre, and a compact $\beta$-type active region (NOAA 12672) near the east limb at N05E61. The corona above the eastern limb included a bright and extended loop system associated with active region AR\,12672, brighter than other structures detected above the western limb, so the east limb was selected as the target. Figure~\ref{figure1} (upper panels) shows the corona above the east limb in the \fexiv\ green-line obtained with our instrument, as well as the corona above the circumference of the Moon's limb in white-light just after the second contact of the eclipse on 21 August 2017 at 17:28:23.64\,UT. The lower-left panel of Figure~\ref{figure1} indicates the signal-to-noise ratio corresponding to this green-line observation, while the AIA 171~\AA~image (taken at 17:26:46\,UT) is shown in the lower-right panel.

%  Fig. 1                                              One column figure
%-----------------------------------------------------------------
%-----------------------------------------------------------------

%--------------------------------------------------------------------
% Section 5
\section{Eclipse Observations}

After delivery to the observing site, the whole system was assembled and adjusted, and the equatorial drive was aligned the night before the eclipse using stars as light sources. The computers and monitors were protected from direct solar illumination before and after totality using the shade of the transportation container, which was also employed as our observing station.

During the days preceding the eclipse, the weather at the observing site was perfect, and forecasts for the eclipse-viewing conditions were good. On the day of the eclipse and during totality, the sky to the north-west of the observing site was partially covered with scattered clouds. In addition, a transparent layer of fairly uniform mist was just perceptible above the observing site towards the Sun.

Observations of the eclipse started just after first contact and continued until fourth contact. A pre-programmed solar tracking rate of the drive was applied. Before and after totality, the Sun was observed through a Baader filter (Mylar foil) to evaluate the long- and short-period stabilities of the telescope, as well as to collect flat-field and dark-current data. The Mylar foil was removed from the aperture of the telescope just after the final Baily's beads disappeared on the east limb, at second contact. Data recorded by both cameras during totality were displayed in real-time, allowing a coarse inspection of the telescope pointing. Exposure parameters selected for the green-line channel during the pre-eclipse trials on the full Moon were applied. The Mylar foil was removed in front of the aperture of the telescope just after the first Baily's bead was seen on the western limb at third contact. Data recorded by both cameras during totality were displayed in real-time, allowing a coarse inspection of the telescope pointing. Exposure parameters selected for the green-line channel during the pre-eclipse trials on the full Moon were applied. The Mylar foil was removed from the telescope aperture just after the first Baily's bead was seen on the western limb at third contact.

As a result, 429 well-exposed images of the solar corona in the green coronal line were recorded during a 122.9-second portion of the totality, starting from 7\,seconds after second contact. According to the eclipse circumstances, calculated using the Solar Eclipse Computer provided by the Astronomical Applications Department of the US Naval Observatory, totality for the observing site was predicted to last 2\,minutes 14.3\,seconds, so our images covered $\approx 92$\,\% of totality. The unprocessed images show a rather diffuse green-line corona over the eastern limb. However, the same images with enhanced contrast revealed many well-defined bright loop-like and pillar-like structures, between other numerous loops of various inclinations, some of them associated with NOAA Active Region 12672 near the east limb visible in SDO/AIA images (Figure~\ref{figure1} lower-right panel). Emission in the \fexiv\ green-line is detected out to a radial distance of about 120,000\,km (166\,arcseconds) above the limb, and to 160,000\,km (220\,arcseconds)  above the limb near the active region (Figure~\ref{figure1}, upper-left panel).

The brightest active-region structures seen in the 2017 images have a signal strength of up to 3000\,DN, while the average coronal structures near the Moon's limb have a signal of $\approx$\,1000\,--\,2000\,DN. An electronic gain of 3.8 was selected for the pre-amplifier of the Andor iXon 885 camera used for the green-line channel, {\em i.e.} each data number corresponds to 3.8 recorded photons. Assuming a Poisson statistical distribution of photon counts, it is therefore possible to search for intensity fluctuations $[\Delta I / I]$ of $0.01$. However, taking into account all instrumental effects, the effective signal-to-noise ratio calculated for recorded data was $\approx 60$ or lower, hence we are able to search for intensity fluctuation $[\Delta I / I]$ of $\approx 0.02-0.03$. This accuracy is still considerably better (by a factor of ten) than for the camera used in our previous eclipse observations. The theoretical spatial resolution of our instrument is defined by the pixel size of the camera rather than the telescope aperture. Although seeing somewhat degraded this resolution, inspection of the image quality revealed that structures of $\approx$\,2\,arcseconds were discernible, with a brief period (37\,seconds to 42\,seconds after second contact) when the seeing was worse. Incidentally, as with our previous observations, it was noticeable that coronal structures crossing each other resulted in enhanced intensities, as expected for optically thin plasmas.

%--------------------------------------------------------------------
% Section 6
\section{Stability and Coalignment of Images}

Although a very reliable equatorial telescope mount was employed, the pointing showed small-scale imperfections,
which consisted of two components. The main one was a slow drift caused by errors in an adjustment of the drive axes, while the second was random sub-pixel and quasi-stochastic variations of the pointing in both axes. As a result, fine adjustments had to be applied to achieve stationary images, so that a photometric analysis could subsequently be applied.

\begin{figure*}
%   \centering
%   \subfigure{\includegraphics[width=9cm]{fig3a.eps}}
%   \subfigure{\includegraphics[width=9cm]{fig3b.eps}}
\centerline{\includegraphics[width=13cm]{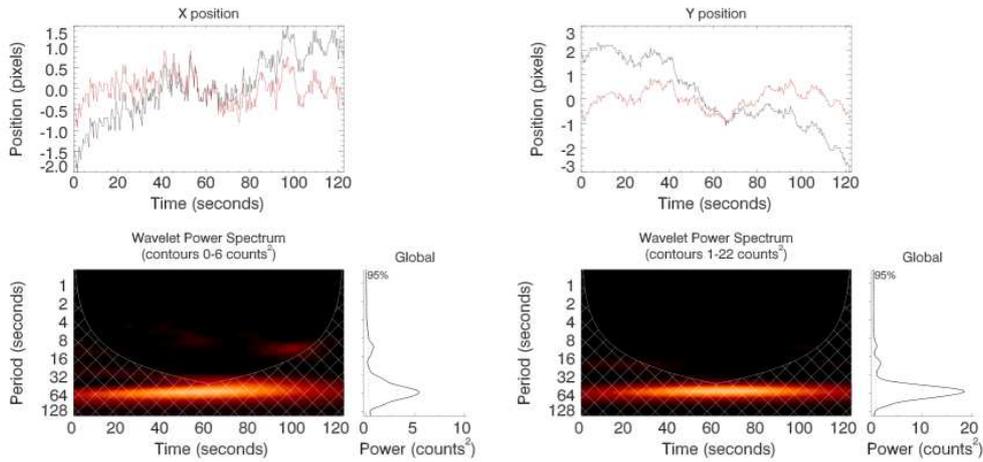}}
%\centerline{\hspace*{0.005\textwidth}
%               \includegraphics[width=0.49\textwidth,clip=,angle=0]{fig3a.eps}
%               \hspace*{0.01\textwidth}
%               \includegraphics[width=0.49\textwidth,clip=]{fig3b.eps}}
      \caption{Temporal variations in the positions of the images collected in the \fexiv\ green coronal line during the totality in the reference system of the CCD sensor. The variations of the positions were calculated using two-dimensional correlations of the observed coronal structures. \textbf{Upper panels:} relative translations with ({\it black lines}) and without ({\it red curves}) a general linear trend. The slightly non-linear drift across the field-of-view was caused by the imperfect adjustment of the equatorial drive's rotation axes. The $X$- and $Y$-directions are along the rows and columns of the sensor respectively. \textbf{Lower panels:} wavelet power spectra of the relative translations without the general linear trend. No high-frequency periodic translations were detected.}
      \label{figure2}
  \end{figure*}

For the exposure time of 250\,ms that we used, the dark-current was $\approx 35$\,DN. After dark-current subtraction and flat-fielding, two different methods for image stabilization were applied. The first is based on an evaluation of momentary positions of the Moon's limb by fitting it with a circle using the automated {\sf MPFITELLIPSE} least-square fitting procedure of C.\,Markwardt, part of the SolarSoftWare (SSW) package. Following this, we numerically shifted the images to positions consistent with the calculated expected one for the Moon with respect to the solar corona in the FoV of the camera. The second method is based on two-dimensional correlations of the sub-images of the bright coronal structures re-binned to ten-times-smaller sub-pixels, {\em i.e.} the resulting shifts have a 0.1\,pixel accuracy. This method gave better stability of the resulting series of images, despite being very sensitive to, for example, the selection of the correlation fields and the image enhancements applied. However, even with the overall fine precision of the stabilization, small sub-pixel jitters of the shifted images persisted, mostly due to the influence of seeing as well as the gradual variation of local brightness due to the advancing Moon's limb. Figure~\ref{figure2} (upper panels) shows the temporal variations in the position of the images collected during the totality in the reference system of the CCD sensor. Variations of the positions were calculated using two-dimensional correlations of the coronal structures seen in the \fexiv\ green line. The slightly non-linear drift across the FoV (shown in the first and third panel from the top) was caused by the imperfect adjustment of the rotation axes of the equatorial drive.

% Fig. 2                                             Two columns figure
%-----------------------------------------------------------------
%-----------------------------------------------------------------

To detect periodic mechanical instabilities in the instrument, which might have affected the photometric analysis of the coronal emission, a wavelet transformation was used to search for periodic variations in the calculated $X$- and $Y$-shifts of the images. A similar procedure was applied in earlier analyses (see \citet{2010SoPh..267..305R} for details). No indications of any instrumental oscillations were found in the frequency range above $0.25$\,Hz. The results of the analysis are shown in Figure~\ref{figure2} (lower panels).

% Fig. 3                                            One column figure
%-----------------------------------------------------------------
   \begin{figure}
\centerline{\includegraphics[width=1.0\textwidth]{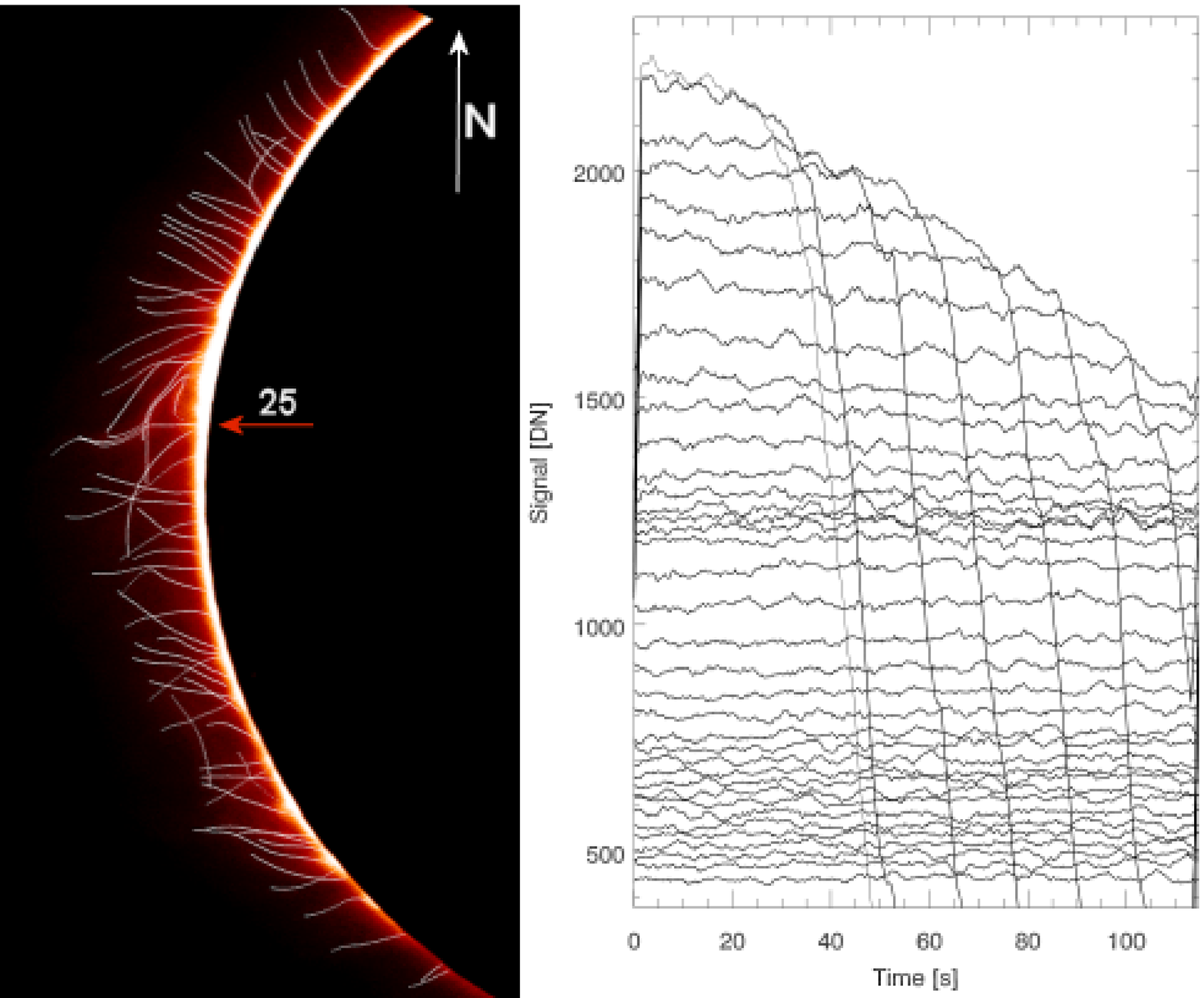}}
      \caption{\textbf{Left panel:} locations of the 65 paths selected to search for any fast-mode magneto-acoustic waves in the loop-like and pillar-like magnetic structures observed above the east limb during eclipse totality (the arrow with N indicates solar North). Selected paths are indicated as chains of white dots in the enhanced-contrast image of the solar corona recorded in the green line. The red arrow shows path number 25 (see text). \textbf{Right panel:} light curves (every third one is shown for clarity) for points along the path. The lowest signals are for the points farthest from the limb. The data are smoothed using a ten-point temporal running average, while the contrast of the coronal image is numerically enhanced. }
      \label{figure3}
  \end{figure}
%-----------------------------------------------------------------

%--------------------------------------------------------------------
% Section 7
\section{Photometric Data Analysis}

% Subsection 7.1
\subsection{Differential Images}

To detect macroscopic displacements of the bright coronal structures viewed against the sky, we analyzed series of differential images. These were calculated as differences between i) consecutive images, ii) consecutive averaged images, and iii) images or averaged images separated by large time-steps. Although the averaged images have a lower temporal resolution, their decreased noise levels enable very small differences in a time sequence to be detected. After a thorough analysis of the differential images, we did not detect any macroscopic displacements of the observed coronal structures during the totality. Some very faint differences, barely visible on some images, are caused by under-corrected instabilities of the pointing rather than real motions of the coronal structures.

% Fig. 4                                              Two columns figure
%-----------------------------------------------------------------
\begin{figure*}
\centerline{\includegraphics[width=0.95\textwidth]{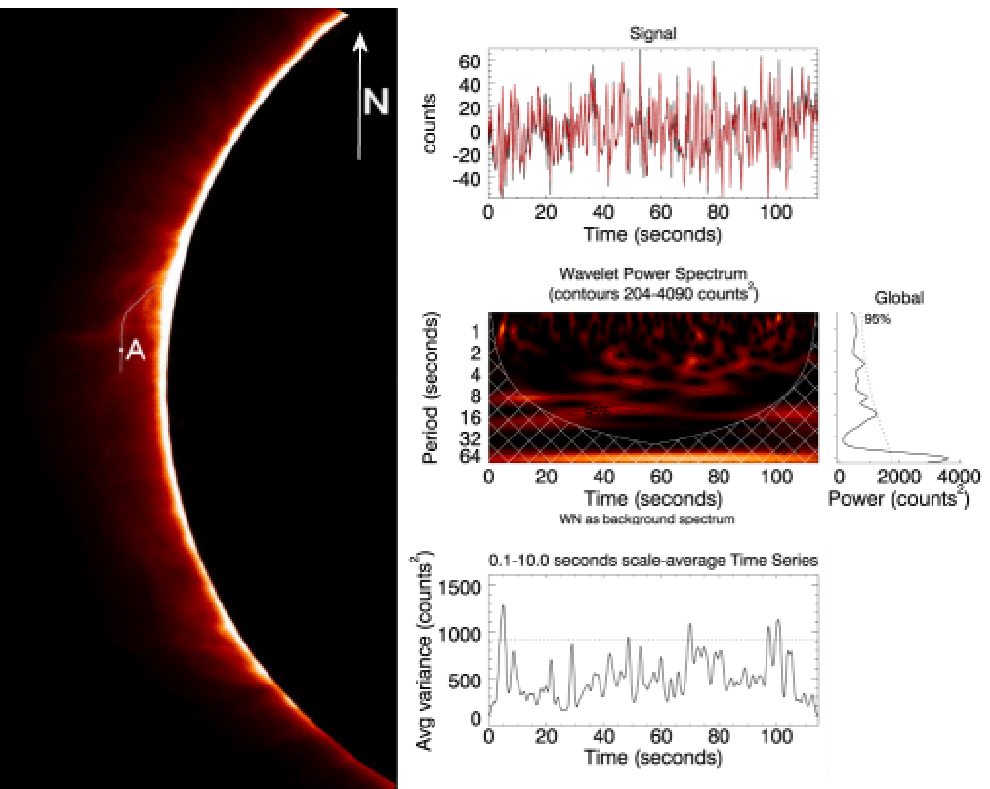}}
\caption{Wavelet analysis of the local brightness variations of the
solar corona recorded in the \fexiv\ green coronal line during the 21
August 2017 eclipse.  \textbf{Left panel:} an enhanced image of the corona with
over-plotted path, and the analyzed pixel marked with a
white dot and labelled A (the arrow with N indicates solar North). \textbf{Right panels:} (upper) variations of the measured signal in point A; (middle left) the wavelet power spectrum with an
over-plotted cone of influence; (middle right) global power {\it vs.} period;
(lower) global power {\it vs.} time. The applied significance limit is 95\,\%.
}
       \label{figure4}
   \end{figure*}
%-----------------------------------------------------------------

% Subsection 7.2
\subsection{Search for Fast-Mode Magneto-Acoustic Waves}

To search for any evidence of fast-mode magneto-acoustic waves travelling along coronal loops in our coronal images, we analyzed variations of the local brightness measured at numerous points selected along the bright structures. The paths were selected manually, following visible coronal structures, although some were chosen that did not do this. The selected paths appear to be representative of all visible structures. Figure~\ref{figure3} (left panel) shows the locations of the 65 paths analyzed. For one path in particular (No.~25 in the analysis), light curves are shown in the right panel of the figure. To enhance the statistical quality of the light curves, the data are smoothed using a ten-point running average. No systematic time shifts between the light curves along any path examined, and no propagating brightness variations along any path, were detected. Magnitudes of the detected local intensity variations, calculated as ratios of a standard deviation to a mean value of the smoothed signal, were of the order of 0.01\,--\,0.015 for all points along all paths.

% Subsection 7.3
\subsection{Wavelet Analysis of the Coronal Emission}

Short-period variations of coronal emission in the green line were searched for in the bright coronal structures, and within the entire FoV of the instrument. To detect short-lived periods of local oscillations, as well as search for a broad spectrum of frequencies, starting from 1.6\,Hz down to 0.0025\,Hz, we applied wavelet transform techniques \citep[see, {\it e.g.,}][for details]{{1998BAMS...79.61T},{1997ApJ...482.1011S}}. The Morlet wavelet was used, and confidence limits assigned to features in the wavelet power spectra that appeared to be significant at 95\,\% and 99\,\% levels as lower thresholds \citep[see][for details]{2010SoPh..267..305R}.

% Subsection 7.3.1
\subsubsection{Wavelet Analysis in Selected Locations}

Wavelet power spectra were calculated for numerous points located along randomly selected paths, following obvious loop-like coronal structures. Figure~\ref{figure4} shows an example of the wavelet analysis of brightness variations in a single pixel (marked A) in path No.~3 in our analysis. The wavelet power spectrum is over-plotted with the cone of influence, {\em i.e.} a region of spectrum where periodicities do not have statistical significance because they are substantial fractions of the time span of the observations. The applied significance limit was 95\,\%. As can be seen from the top panel (time variations of the signal in point A), no obvious periodicity is evident, while the plot of global power against period more clearly shows that any possible periodicity has a less than 95\,\% significance.

Wavelet analyses of a large number of other points within coronal structures such as the one illustrated in Figure~\ref{figure4} typically resulted in insignificant power for the periodicities examined. However,  many short-lived increases of the power spectra above the selected threshold level of 95\,\% were detected in the upper parts of the visible structures and indeed parts of the corona where there were no evident structures. These always had low signal levels, and it is clear that the apparent rapid variations can invariably be attributed to noise.

Wavelet analysis is resistant to variations in the mean signal level. However, to be certain that the results obtained are also correct in lower parts of the paths, where the coronal structures were gradually covered by the encroaching limb of the Moon, additional power spectra were calculated for data ``clipped'' to the period defined by when the signals were unaffected by the advancing limb. The clipped time series become shorter, but with a uniform signal level. Once again, we found no obvious evidence of periodicities within coronal structures in these regions.

% Subsection 7.3.2
\subsubsection{Wavelet Analysis for All Points Inside the Complete FoV (Comprehensive Analysis)}

From the foregoing, wavelet analysis shows that any apparent periodicities from locations along coronal features have only marginal significance. We made an additional search for periodic behaviour in pixels regardless of whether they are along specific coronal structures. Some 404,000 pixels were examined in this way using 429 intensity measurements, one for each image taken. Wavelet spectra were calculated for 71 frequencies from 1.6\,Hz to 0.01\,Hz, but our interest was focused on the range 1.6\,--\,0.5\,Hz, since waves with periods shorter than 2 seconds may have significance for coronal heating processes.

Results of the large-scale wavelet analysis can be visualised as two-dimensional images (henceforth referred to as maps) showing pixels for which the wavelet power spectra exceeded an assumed significance level (95\,\%) for a particular frequency and time span. A preliminary inspection of the maps showed only a random scatter of points, with no regular patterns or clusters, consistent with noise.

%  Fig. 5                                              One column figure
%-----------------------------------------------------------------
   \begin{figure}[h]
   \centerline{\includegraphics[width=9cm]{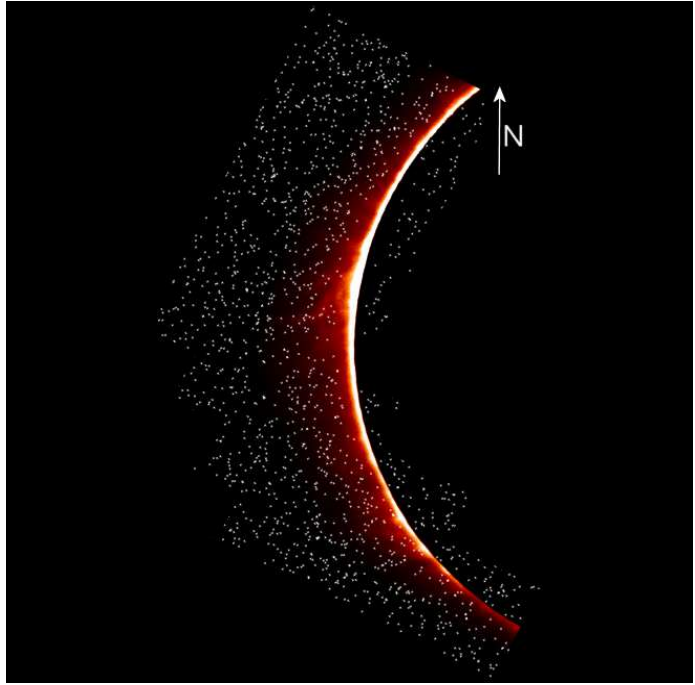}}
      \caption{The distribution of pixels (small white dots) that wavelet analysis indicates have periodicities with a 95\,\% significance level. The points show no alignment with coronal structures, as can be seen by comparison with Figure~\ref{figure3}. This particular analysis was done for images 300 to 336 inclusive, a time span of ten seconds, starting at 84\,seconds after the commencement of the observations. The arrow with N indicates solar North. The Moon's motion relative to the corona is from bottom right to top left, so that the coronal structures on the east limb are seen to best advantage. }
      \label{figure5}
  \end{figure}
%-----------------------------------------------------------------

A more detailed analysis of the maps involves a four-dimensional data cube, with dimensions of two space, one time, and one frequency, with a practically unlimited number of time spans. To narrow the search and make the problem more tractable, we developed an automatic code that inspected time-series of power spectra for all pixels, searching for detections at a 95\,\% significance level over time spans of various lengths and starting points. We considered time spans of at least ten seconds (36 images), taking into consideration that a series of oscillations would be necessary for sufficient energy to be deposited in a coronal structure. As a result of this analysis, we obtained a set of maps with points indicating pixels that show periodicities with $\geqslant 95$\,\% significance. As an example, Figure~\ref{figure5} shows such a map starting 84\,seconds after the commencement of the observations and including a 10.6-second time span (images 300 to 336). There are 5076 points (white dots in the figure) for which the wavelet analysis indicates periodicities with $\geqslant 95$\,\% significance. Their random distribution, without any alignment to particular coronal structures, shows clearly that they have no physical significance. Indeed, some are even present for pixels inside the Moon's disk where there is no coronal signal at all. All maps were examined by eye in the same way but only random distributions of apparent detections were found, uncorrelated with coronal structures. Any oscillations propagating along magnetic-loop structures due to magneto-acoustic waves need not have a single frequency\,--\,two or three may be present depending on the nature of the waves' initiation. To take account of this, we also searched for such oscillations, finding that most of the detections occurred in pixels with low signal levels in the outer corona or even within the Moon's disk, again clearly indicating that they are due to noise.

In summary, based on this extensive wavelet analysis, it appears that, in the two-minute duration of our observations and in the portion of the corona that we viewed, there are no apparent periodicities of any significance in the intensity data with frequencies in the range 0.5\,--\,1.6\,Hz (periods 2\,--\,0.625\,second).

%--------------------------------------------------------------------
% Section 8
\section{Discussion and Conclusions}

Our observations of a portion of the green-line corona during the total eclipse of 21 August 2017 reveal no evidence of intensity oscillations in coronal structures that might be indicators of short-period magneto-acoustic waves. These data were taken with a CCD camera with 16-bit intensity resolution, giving a sensitivity of $\Delta I / I \approx 0.02 - 0.03$ and a spatial resolution of $\approx$ 2\,arcseconds. The 2017 observations covered the corona over the Sun's east limb which included an active region (NOAA 12672); the level of solar activity was low, with a background GOES X-ray level of B3. Similar wavelet and classical Fourier analysis previously used by us for the 1999 and 2001 eclipses led to similar conclusions that there were insignificant intensity oscillations. However, the analysis of \cite{2001MNRAS.326..428W} indicated the presence of a travelling wave of intensity oscillation in the active region seen in the 1999 eclipse. Although of marginal significance, the wave appears to be sustained but with decreasing amplitude over an active-region loop length, with a period of $\approx$\,six\,seconds. This was interpreted by \cite{2003A&A...409..325C} as being a fast-mode magneto-acoustic wave with the amplitude modified by line-of-sight effects. In the present analysis, as well as examining 404,000 locations, specific searches were made for travelling waves in discernible loop structures, in particular the most prominent one in the green line images illustrated in Figure~\ref{figure3}. No such evidence was found. In attempting to understand this, a difference in the 1999 and 2017 eclipse observations is that the active region (NOAA 8656) seen in 1999 and analysed by \cite{2001MNRAS.326..428W} was stronger and flare-prolific, while NOAA 12672 was weaker with little flare association.

Similar considerations may account for the fact that other observers making similar eclipse observations have found some evidence for periodic intensity fluctuations, in particular those of \cite{1999SoPh..188...89C} and \cite{2009SoPh..260..125S}. \cite{1999SoPh..188...89C} reporting on 1998 eclipse observations when solar activity was high, and with a wavelet analysis similar to ours, found intensity oscillations having multiple periods, the shortest being 6.9\,seconds, and with amplitudes $\Delta I / I = 0.5 - 3.5$\,\%. \cite{2009SoPh..260..125S} observed the 2006 eclipse, which was at a time of low solar activity, but nevertheless they obtained data for a strong active region on the limb, finding periods of $\approx 27$\,seconds in their green-line channel and a shorter period (20\,seconds) in their red-line channel with intensity fluctuations of $\gtrsim 2$\,\% on the boundary of the active region. However, these oscillations have periods substantially longer than the high-frequency MHD waves capable of effectively transporting energy into the solar corona (frequencies > 0.5\,Hz).

The relatively low temporal cadence of spacecraft images of the corona does not allow searches for localized brightness variations with $\approx 1$\,Hz frequencies, attributable to short-period MHD waves \citep{1994ApJ...435..482P}. Possibly observations with CCD cameras mounted on large ground-based coronagraphs under excellent atmospheric conditions will in time be used to look for such brightness variations. Meanwhile, instruments with high-quality CCD cameras and telescopes employed during total solar eclipses, such as the instrument described here, offer the best means of making high temporal resolution observations needed to search for MHD waves involved in the coronal-heating process. The observations reported here suggest that intensity oscillations and travelling waves, at least at low solar-activity levels, are at best extremely subtle. This should help in planning for future total eclipses and, taking a cue from the CoMP coronagraph observations of \cite{2007Sci...317.1192T}, a search for velocity fluctuations using observations of the profile of the coronal green or red lines in combinations with intensity observations could be more revealing. We are at present investigating this as a means of looking for short-period wave motions during solar eclipses, and in due course we will be reporting progress and results.

% Acknowledgements
\begin{acks}
A.\,Berlicki, K.\,Radziszewski, and P.\,Rudawy were supported by the National Science Centre in Poland, under grant No.\,UMO-2015/17/B/ST9/02073. D.\,B. Jess and F.\,P. Keenan are grateful to the UKRI Science \& Technology Facilities Council for financial support under grant ST/P000304/1, and P.\,H. Keys is grateful to the Leverhulme Trust for the award of an Early Career Fellowship. A.\,Berlicki acknowledges support from the Czech Science Foundation (GACR) through the grant No.\,16-18495S and from the project RVO:67985815 of the Astronomical Institute of the Czech Academy of Sciences.
\end{acks}

% Disclosure
\section*{Disclosure of Potential Conflicts of Interest}

The authors declare that they have no conflicts of interest.

% for the bibliography, at the end
\bibliographystyle{spr-mp-sola} % style spr-mp-sola.bst
\bibliography{rudawy_et_al_2018} % your references rudawy_et_al_2018.bib
%
%-----------------------------------------------------------------------------------
%
\end{article}

\end{document}